\documentclass[a4paper,fleqn]{cas-dc}

\usepackage[numbers,sort&compress]{natbib}

\usepackage{adjustbox}
\usepackage{float}
\usepackage[ruled]{algorithm2e}
\usepackage{algorithmic} 
\usepackage{CJKutf8}
\usepackage{enumitem}
\usepackage{multirow}
\usepackage{graphicx}
\usepackage{color}
\usepackage[normalem]{ulem}
\useunder{\uline}{\ul}{}

\def\tsc#1{\csdef{#1}{\textsc{\lowercase{#1}}\xspace}}
\tsc{WGM}
\tsc{QE}
\tsc{EP}
\tsc{PMS}
\tsc{BEC}
\tsc{DE}

\usepackage{caption}
\begin{document}
\begin{sloppypar}
\let\printorcid\relax
\let\WriteBookmarks\relax
\def\floatpagepagefraction{1}
\def\textpagefraction{.001}
\begin{CJK}{UTF8}{gbsn}

\shorttitle{}

\shortauthors{J.Zhou et~al.}

\title [mode = title]{CPLOYO: A Pulmonary Nodule Detection Model with Multi-Scale Feature Fusion and Nonlinear Feature Learning}

\author[1]{Meng Wang}
\ead{wangmeng123@tongji.edu.cn}

\author[2]{Zi Yang}
\ead{hxkb122@tzc.edu.cn}

\author[1,3]{Ruifeng Zhao}
\ead{qglxhsyz@hhu.edu.cn}
\cormark[1]

\author[4]{Yaoting Jiang}
\ead{yaotingj@alumni.cmu.edu}

\address[1]{Department of Radiotherapy, Tongji Hospital, School of Medicine, Tongji University , Shanghai, 200065, China}
\address[2]{Department of Nuclear Medicine, Shanghai Pulmonology Hospital, School of Medicine, Tongji University , Shanghai, 200065, China}
\address[3]{Department of Radiotherapy, Shanghai Pulmonology Hospital, School of Medicine, Tongji University, Shanghai, 200065, China}
\address[4]{Carnegie Mellon University, USA}

\cortext[cor1]{qglxhsyz@hhu.edu.cn}

\begin{abstract}
The integration of Internet of Things (IoT) technology in pulmonary nodule detection significantly enhances the intelligence and real-time capabilities of the detection system. Currently, lung nodule detection primarily focuses on the identification of solid nodules, but different types of lung nodules correspond to various forms of lung cancer. Multi-type detection contributes to improving the overall lung cancer detection rate and enhancing the cure rate. To achieve high sensitivity in nodule detection, targeted improvements were made to the YOLOv8 model. Firstly, the C2f\_RepViTCAMF module was introduced to augment the C2f module in the backbone, thereby enhancing detection accuracy for small lung nodules and achieving a lightweight model design. Secondly, the MSCAF module was incorporated to reconstruct the feature fusion section of the model, improving detection accuracy for lung nodules of varying scales. Furthermore, the KAN network was integrated into the model. By leveraging the KAN network's powerful nonlinear feature learning capability, detection accuracy for small lung nodules was further improved, and the model's generalization ability was enhanced. Tests conducted on the LUNA16 dataset demonstrate that the improved model outperforms the original model as well as other mainstream models such as YOLOv9 and RT-DETR across various evaluation metrics.

\end{abstract}




\maketitle

\section{Introduction}

Lung cancer is the leading cause of mortality from malignant neoplasms worldwide. Despite significant advancements in medical technology, the mortality rate remains high, exerting considerable pressure on global public health systems. Early detection and diagnosis are critical, as timely intervention can significantly improve survival rates \cite{lyu2018using,wan2021computer,dematteo2024headaches}. Pulmonary nodules often serve as the first indication of lung cancer. These nodules are typically asymptomatic and can be easily overlooked, yet early detection allows for intervention that greatly enhances the chances of successful treatment. Pulmonary nodules usually appear as circular or oval shadows, ranging from 3 to 30 millimeters in diameter on CT scans. Their variation in shape and size presents a challenge for clinicians, particularly when the nodules are small or irregularly shaped \cite{kim2020role,tlijani2023optimized}. Currently, CT scans are the most widely used non-invasive diagnostic tool for screening pulmonary nodules. However, the accurate detection of small and early nodules remains a significant challenge due to the variability in nodule characteristics and the complexity of the pulmonary environment.

\begin{figure*}
    \centering
    \includegraphics[width=1\linewidth]{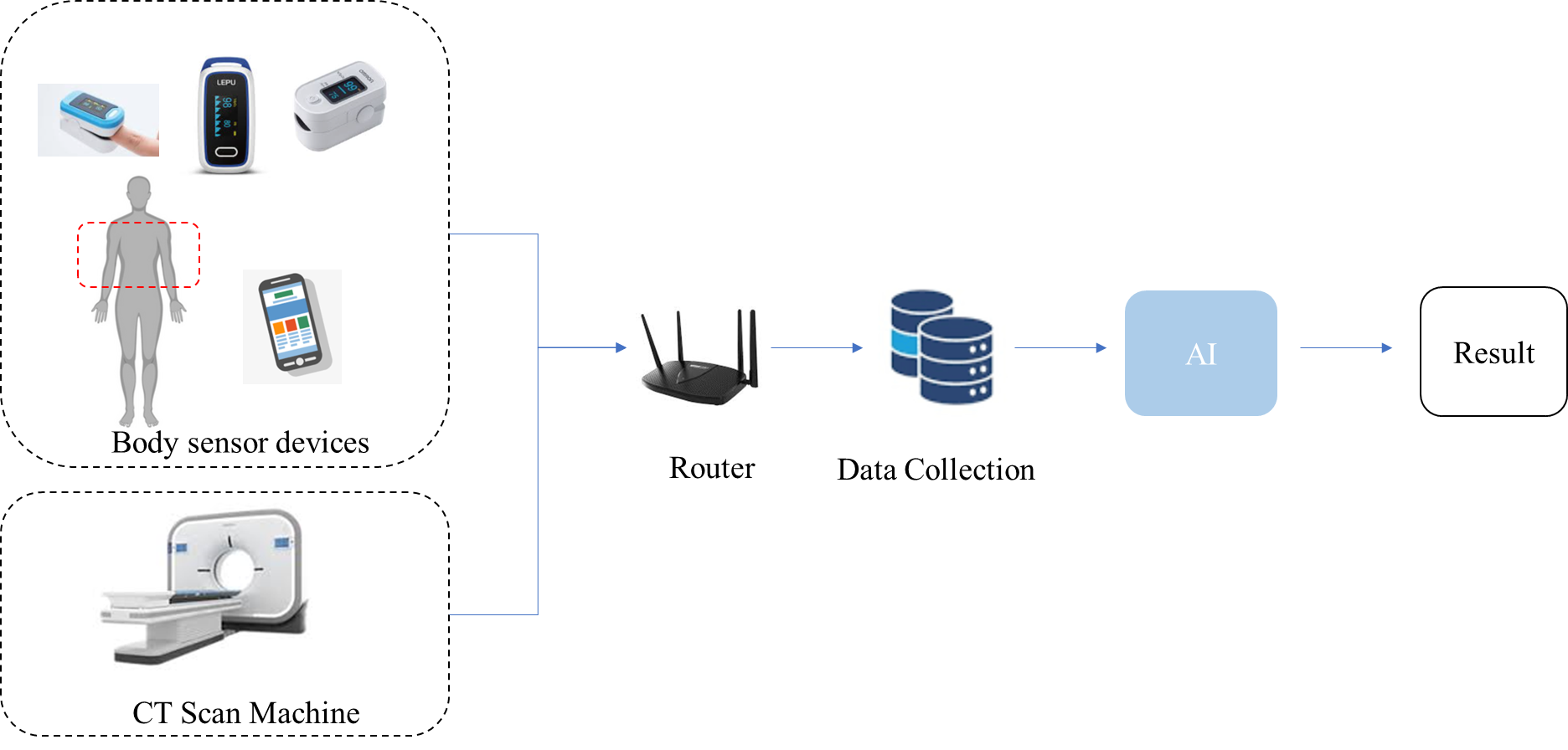}
    \caption{Overview structure of the lung nodule detection and classification system.}
    \label{IOT}
\end{figure*}

Recent advances in the Internet of Things (IoT)~\cite{wang2024hybrid,xu2024enhancing,krieger2020system} offer promising opportunities to address the limitations in early lung cancer detection. IoT-enabled devices~\cite{pang2024electronic,wang2024using,lyu2024optimized,dong2024design,huang2024risk,wang2024cross,jiang2020dualvd} allow for continuous patient monitoring, real-time data collection, and seamless integration of diagnostic tools, thereby improving the overall detection and management process, as illustrated in Figure \ref{IOT}. For instance, wearable sensors can track patients' respiratory patterns and vital signs, providing valuable data that can be analyzed in conjunction with CT scans to identify early signs of lung abnormalities. Moreover, IoT-connected CT scanners can automatically upload imaging data to cloud-based platforms, where advanced object detection algorithms, such as those based on deep learning, can process the information in real-time. This continuous data flow ensures timely access to critical diagnostic information, facilitating early intervention.

In recent years, object detection techniques have garnered significant attention, particularly in the field of automated medical diagnostics. Computer vision technologies employ object detection algorithms to automatically analyze medical imaging data, efficiently identifying and localizing suspicious regions, which enhances diagnostic accuracy and efficiency~\cite{ren2025iot,li2024ltpnet,liu2025real,xu2022dpmpc,wang2024intelligent,zhou2024optimization,lee2024traffic}. These modern approaches outperform traditional methods, which rely on manual annotation and the experience of clinicians. By improving diagnostic accuracy, they enable more precise treatment decisions. Deep learning-driven object detection algorithms have proven to be highly effective in addressing many of the challenges in medical imaging. Two-stage detection algorithms, such as Region Proposal Networks (RPN) followed by a refinement step, have been the standard approach in this domain \cite{sun2021attention,almayyan2024detection,RenBaoPing2024NewProductiveForces,jin2025rankflow,zhao2025short,an2023runtime,guo2024construction,liu2024dsem,luo2025intelligent,wang2024deep}. However, these two-stage algorithms are plagued by a high false positive rate, especially when dealing with complex backgrounds or variable nodule shapes \cite{zhao2018lung,wang2025physically}. This results in an excessive number of false-positive detections, decreasing diagnostic efficiency. Additionally, these methods demand substantial computational resources and time, particularly when processing high-resolution medical images \cite{liu2020automatic,zhuang2020music,chen2024enhancing,richardson2024reinforcement,tang2025real,wang2024recording}. While two-stage algorithms exhibit strong accuracy in certain situations, they face significant challenges in clinical applications.

The integration of IoT~\cite{Wang2024Theoretical,peng2025integrating,sui2024application,peng2024automatic,yan2024application} is pivotal in overcoming the limitations of traditional two-stage detection algorithms. IoT devices can collect vast amounts of patient data, including data from wearable devices, electronic health records, and environmental sensors. This rich dataset allows for more comprehensive training of detection algorithms, thereby improving their ability to differentiate between benign and malignant nodules in complex backgrounds. IoT also enables distributed computing resources that facilitate real-time processing and analysis of high-resolution images. This eliminates the need for centralized, resource-heavy infrastructure and accelerates the detection process while enhancing scalability, making advanced diagnostic tools more accessible in diverse clinical settings.

To address the shortcomings of two-stage detection algorithms, researchers have developed one-stage detection algorithms \cite{yan2018deeplesion,subramanian2023mdho,liu2025eitnet,zheng2024triz,wan2024image,xi2024enhancing,yuan2025gta,zhang2024deep}. These algorithms combine the detection and false positive suppression steps into a unified model, achieving faster detection speeds. One-stage algorithms directly predict nodule categories and positions using deep neural networks, eliminating the proposal generation and post-processing steps characteristic of traditional two-stage methods. This results in reduced computation time and resource consumption. Furthermore, one-stage algorithms offer high accuracy, making them well-suited for applications that require real-time processing or large-scale data screening. However, one-stage models still face challenges, particularly when detecting small nodules or dealing with interference from high background noise. These algorithms often miss or misdetect smaller lung nodules or those obscured by other tissues, especially when the nodules are irregularly shaped or small. Enhancing the sensitivity of one-stage algorithms to small nodules while minimizing false positives remains a significant challenge in this field.

Integrating IoT into one-stage detection frameworks is crucial for enhancing model performance by providing real-time data streams and supporting continuous learning. IoT infrastructure enables the real-time collection and transmission of imaging data, which can be immediately processed by one-stage algorithms running on edge devices or cloud platforms. Real-time processing ensures that detection models are continuously updated with the latest patient data, improving sensitivity to small nodules and reducing false positives. Additionally, IoT enables the seamless integration of environmental and lifestyle data, providing valuable context to further refine the accuracy of nodule detection algorithms.

YOLOv5 is a leading one-stage object detection algorithm, renowned for its computational efficiency and exceptional performance across various computer vision tasks. It is particularly adept at processing spatial information and has demonstrated remarkable results in object detection. YOLOv5's adaptive anchor box computation and enhanced feature fusion mechanisms improve both the speed and accuracy of nodule detection \cite{kim2023application}. Its multi-scale feature fusion strengthens the model's ability to detect objects of various sizes, allowing it to perform well in complex backgrounds. Despite its strong performance, YOLOv5 still faces limitations when dealing with complex backgrounds and small targets, such as pulmonary nodules. The small size, diverse shapes, and frequent occlusions of pulmonary nodules—especially those located in dense anatomical regions—pose challenges to YOLOv5's feature extraction and localization capabilities. While YOLOv5 excels in speed, its accuracy and robustness require further improvement for fine-grained medical imaging tasks. As a result, researchers are exploring the integration of advanced technologies, such as Transformers, to overcome YOLOv5's shortcomings in capturing fine details and recognizing small targets.

In this paper, we introduce CPLOYO, a novel object detection method designed to rapidly and accurately detect small pulmonary nodules in medical imaging. CPLOYO’s architecture integrates advanced techniques to enhance both detection accuracy and efficiency. It features a Pixel-Wise Spatial Attention (PSA) mechanism in the backbone to improve pixel-level feature extraction, along with the Convolutional Block Attention Module (CBAM) to enhance feature fusion in the Neck. The model introduces a lightweight yet powerful detection module, C2f\_RepViTCAMF, which combines the advantages of RepViT and Contextual Attention with Multi-scale Feature Fusion (CAMF) to effectively detect small and complex objects like pulmonary nodules. Additionally, CPLOYO employs a Feature Pyramid Network (FPN) and a Path Aggregation Network (PAN) alongside CBAM modules. A key innovation of CPLOYO is the KAN-Bottleneck(Kolmogorov-Arnold Networks-Bottleneck) layer, which replaces traditional multi-layer perceptron (MLP) structures with a novel, learning-based activation mechanism. This innovation ensures optimal network performance for even the most challenging detection tasks.

\begin{enumerate}
    \item \textbf{Enhanced Feature Extraction with PSA and CBAM}: By integrating \textbf{Pixel-wise Spatial Attention (PSA)} and \textbf{Convolutional Block Attention Module (CBAM)}, CPLOYO strengthens the feature extraction process at multiple levels—both spatial and channel, improving the detection of small and complex objects such as pulmonary nodules.
    
    \item \textbf{Lightweight and Fast Detection with C2f\_RepViTCAMF}: The introduction of the \textbf{RepViT} module, optimized for efficiency with structural reparameterization and multi-scale context capturing, enhances the model's ability to detect small pulmonary nodules while maintaining fast training and inference speeds.
    
    \item \textbf{Innovative KAN-Bottleneck for Improved Expressive Power}: The \textbf{KAN-Bottleneck} replaces traditional MLP layers with flexible, learnable activation functions, enabling the network to capture more complex relationships in high-dimensional data, thus improving detection accuracy and performance in complex tasks.
\end{enumerate}

\section{Related work}

\subsection{Traditional Machine Learning}

One of the key challenges in lung nodule detection algorithms is the accurate localization of the nodule's position. Traditional methods often involve manual feature extraction, followed by the input of these features into a classifier for the detection and classification of lung nodules in CT images. Croisille et al. demonstrated that extracting pulmonary vessels from CT images and performing automatic segmentation could significantly improve the detection of lung nodules \cite{croisille1995pulmonary}. Kostis employed a three-dimensional approach for lung nodule detection in CT imaging, discussing isotropic resampling techniques for anisotropic CT data and segmentation algorithms based on three-dimensional intensity and morphological features. He developed a new model for characterizing volumetric growth in longitudinal CT studies \cite{kostis2003three}. Massoptier et al. used lung parenchyma masks generated through thresholding and morphological techniques, although they still encountered challenges related to over-segmentation and under-segmentation in some images \cite{massoptier2009automatic}. Messay and his team utilized various lung nodule feature information, combining intensity thresholds and morphological processing to achieve good results in lung nodule detection \cite{messay2010new}. Kumar et al. proposed a two-dimensional Otsu algorithm based on Darwinian Particle Swarm Optimization (DPSO) and Fractional-Order Darwinian Particle Swarm Optimization (FODPSO) for segmenting lung parenchyma from CT images \cite{kumar2019modified}.

Additionally, traditional machine learning methods have also been widely applied in lung nodule detection. Ye et al. performed lung parenchyma segmentation using fuzzy thresholding, followed by the enhancement of object representation for lung nodules using volume shape index maps and point maps. A rule-based method was then applied to eliminate easily removable nodule candidates, and Support Vector Machines (SVM) were used for classification, reducing the number of false positives \cite{ye2009shape}. Zhang Jing et al. introduced SVM and rule-based methods for lung nodule recognition. In the network operations, irrelevant candidate targets were first excluded, potential lung nodule targets were then filtered out, and SVM was used for target classification, leading to the final recognition results. Khan et al.\cite{khan2019lungs} employed contrast enhancement, segmentation, and feature extraction processes on the images, followed by training and testing the extracted features using a pre-selected classifier. Experimental results showed that this method was highly effective in reducing false positive rates (FPRs) and demonstrated high sensitivity. Messay et al. combined intensity thresholds and morphological processing to detect and segment candidate nodules, and used sequential forward selection to determine the optimal feature subset for two different classifiers, achieving a higher detection rate. Lee et al. utilized Fisher Linear Discriminant Analysis (LDA) for the final classification of nodules; however, this approach may lead to a decrease in the generalization performance of the network \cite{lee2010computer}. Murphy and his team proposed an automatic detection scheme for nodules in chest CT scans and conducted extensive evaluations. They combined shape and local image features of candidate structures in the lung volume with two consecutive K-Nearest Neighbors (K-NN) classifiers to reduce the false positive rate \cite{murphy2009large}. Toğaçar et al. achieved the highest classification accuracy using a combination of AlexNet and K-NN classifiers. They applied the Minimum Redundancy Maximum Relevance (mRMR) feature selection method on deep feature sets to select the most effective features, which were then used in K-NN for classification, resulting in improved accuracy \cite{tougaccar2020detection}. Hawkins et al. achieved the best model with a random forest classifier using 23 stable features, outperforming the Lung Image Database Consortium (LIDC) and morphological methods \cite{hawkins2016predicting}. Lee and his team proposed an innovative cluster-assisted ensemble classification method, utilizing the Random Forest algorithm and clustering-assisted construction of a hybrid random forest-based lung nodule classification structure to achieve superior classification performance \cite{lee2010random}.

Although the traditional pulmonary nodules detection methods have made some achievements in meeting the requirements of computer-aided diagnostic system for the detection effect, these methods have obvious defects because they are based on empirical design characteristics. Therefore, in order to improve the accuracy of pulmonary nodules detection, the detection model must be optimized.

\subsection{Deep Learning Based Method}
With the development of deep learning technologies, particularly the advancements in Convolutional Neural Networks (CNNs), a more efficient approach for lung nodule detection has emerged. CNNs are capable of automatically learning key features during the training process, addressing the limitations of manually designed features, and enhancing the feature extraction capabilities of the network\cite{zhang2025self,wang2025unified,zhang2025cross}. Currently, deep learning-based lung nodule detection algorithms can be categorized into two main types. One category employs a single-stage algorithm, which directly classifies and regresses the input data through the network model, utilizing the concept of anchor boxes for lung nodule detection. The other category uses a two-stage detection algorithm, where candidate regions are generated first and then detected. Lo et al. developed a dual-matching method and artificial visual neural network technology for lung nodule detection. They first used a spherical template dual-matching technique for high-sensitivity initial tuberculosis search of circular objects, and the artificial CNN served as the final classifier to determine whether suspicious images contained lung nodules, with the total processing time of the method being approximately 15 seconds \cite{lo1995artificial}. U-Net has gained significant popularity due to its outstanding performance in medical image segmentation. Its architecture is based on fully convolutional networks, but with an optimization through the introduction of skip connections \cite{ronneberger2015u}. These skip connections allow the network to merge low-level and high-level features at different layers, helping retain more spatial information. Zhou et al. proposed a more robust medical image segmentation architecture, U-Net++, which introduced nested dense skip connections, enabling tighter collaboration between encoder and decoder sub-networks and providing superior performance for medical image segmentation tasks \cite{zhou2018unet++}. Zhang et al. proposed a lung nodule detection algorithm based on attention mechanisms and feature pyramids. By using the ResNet backbone network combined with channel-space attention mechanisms, the network extracts more semantic and positional information. In the prediction stage, a feature pyramid network is used to fuse multi-scale features, improving detection performance for small nodules and nodules near blood vessels. Given that lung nodules are typically small in size, usually ranging from 3-6mm, image segmentation can face challenges such as missed detection and blurry boundaries. Consequently, many experts and researchers in the field of lung nodule detection have been continuously working to address these issues.

Faster R-CNN employs a two-stage detection method \cite{ren2015faster}. First, the Region Proposal Network (RPN) extracts candidate bounding boxes, which are then mapped to fixed-size feature maps using Region of Interest (RoI) pooling. Finally, these features are processed by a classifier and a bounding box regressor to perform object classification and location refinement. By incorporating the RPN and adopting a two-stage detection strategy, Faster R-CNN significantly improves the accuracy and efficiency of object detection, making it one of the classic models in the field. El et al. \cite{el2020two} used Faster R-CNN and SSD in the second stage with Inception-V2 as the backbone, achieving a sensitivity of 96.4\% on the LUNA16 dataset. Tong et al. introduced an iterative self-organizing data analysis technique into the Faster R-CNN model, reducing false positives with a strategy that utilizes a 3D convolutional neural network to exploit the 3D nature of CT images and addressing class imbalance using focal loss \cite{tong2020pulmonary}. Setio et al. proposed a multi-view convolutional network, combining candidates obtained from three specialized nodule detectors designed for solid, sub-solid, and large nodules. The network achieved high detection sensitivities of 85.4\% and 90.1\% with 1 and 4 false positives per scan, respectively \cite{setio2016pulmonary}.

Compared to traditional 2D methods, 3D convolutional neural networks (CNNs) leverage their hierarchical structure designed for 3D samples, which allows them to capture rich spatial information more effectively. This enhances the model’s ability to perceive and understand the data, thereby acquiring more representative features. In addressing the challenges posed by the vast variation and complex simulations of lung nodules, Dou et al. proposed a simple yet effective strategy for capturing multi-level contextual information \cite{dou2016multilevel}. Zhu et al.\cite{zhu2018deeplung} considering the 3D nature of CT data and the characteristics of dual-path networks, proposed two deep 3D dual-path networks for nodule detection and classification. Validation on the LIDC-IDRI dataset demonstrated that their method achieved diagnostic performance comparable to that of experienced physicians, both at the nodule and patient levels. Zhao proposed a 3D CNN-based lung nodule detection algorithm with a multi-scale attention mechanism that explores the relationships between features from both spatial and channel perspectives. This strengthens the features, making it more conducive to object localization and bounding box regression.

The accuracy and reliability of lung nodule detection are largely influenced by data processing methods, especially during the image preprocessing and feature extraction stages. Worku J introduced a "denoising-first" dual-path convolutional neural network (CNN) for CT image denoising, which achieved promising results in experiments \cite{sori2021dfd}. Wang et al. conducted an in-depth study of the core components of Super-Resolution Generative Adversarial Networks (SRGAN), including the network architecture, adversarial loss, and perceptual loss, and proposed improvements to these components. They introduced the Enhanced Super-Resolution Generative Adversarial Networks (ESRGAN), designed to achieve higher visual quality, making generated images appear more realistic and natural in texture \cite{wang2018esrgan}. Convolutional neural networks have made significant advancements in medical image processing, providing efficient tools for disease diagnosis and analysis.

\section{Method}
\begin{figure*}
    \centering
    \includegraphics[width=0.7\linewidth]{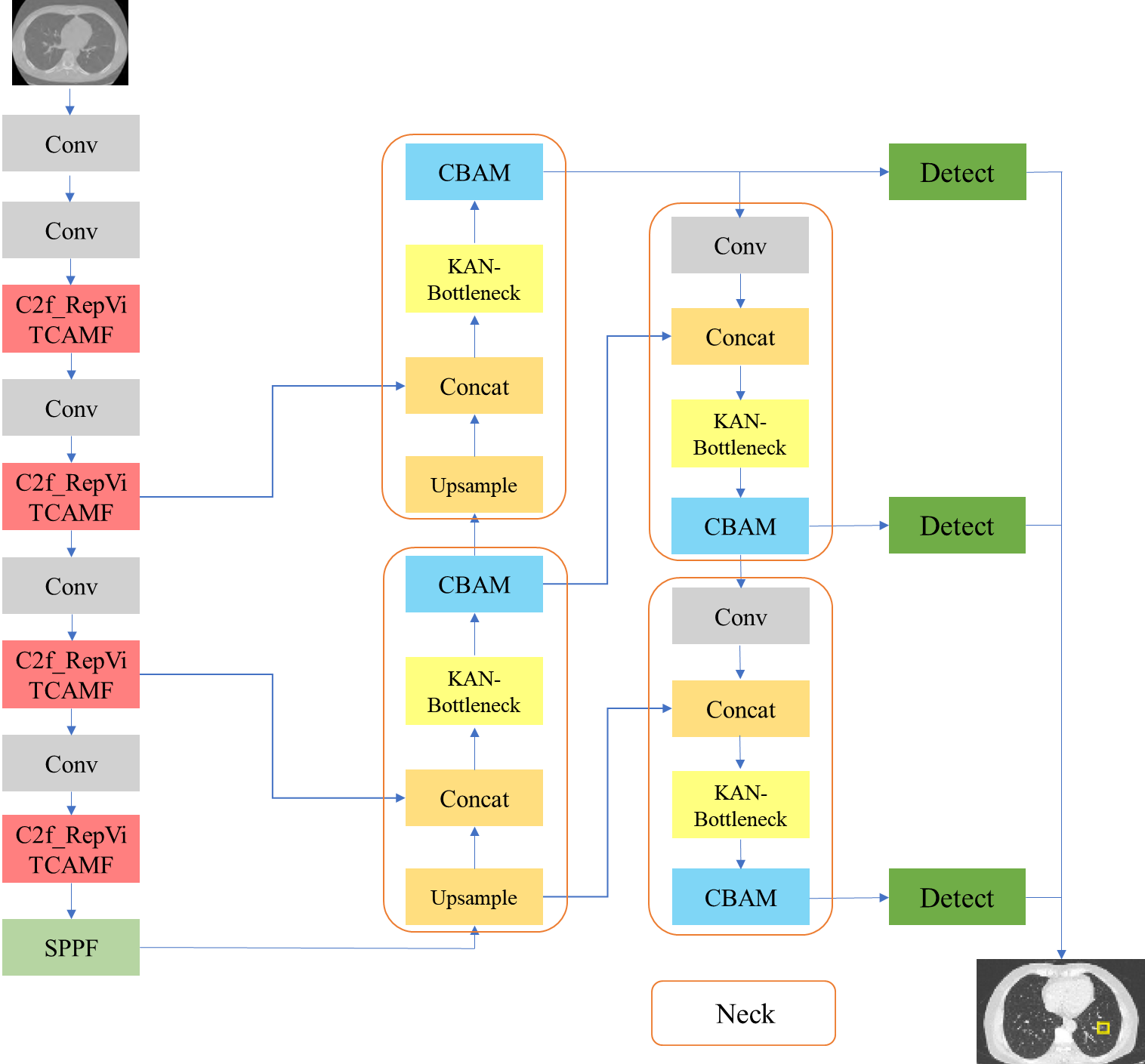}
    \caption{Overall network architecture}
    \label{overall_fram}
\end{figure*}

We propose an object detector named CPLOYO. In the Backbone, we incorporate the Pixel-wise Spatial Attention (PSA) mechanism, which enhances the model's ability to extract object features during training, thereby improving the performance of object detection. In the Neck, we introduce the Convolutional Block Attention Module (CBAM), which further improves feature fusion and information propagation, thereby boosting the overall performance of the model. Figure \ref{overall_fram} illustrates the CPLOYO architecture. The C2f\_RepViTCAMF module improves the model's ability to detect small pulmonary nodules by combining the RepViT architecture with multi-scale context enhancement, using depthwise separable convolutions to reduce computational complexity and enhancing the fusion of local and global information through multi-scale context capturing. The CPYOLO Neck module integrates a Feature Pyramid Network (FPN) and KAN Bottleneck module for effective multi-scale feature extraction, optimizing the feature learning process, while the CBAM module further strengthens the model's focus on key information through channel and spatial attention mechanisms. The innovative design of these two modules significantly enhances the model's accuracy and efficiency in detecting small objects in medical imaging.

\subsection{C2f\_RepViTCAMF}
\begin{figure}
    \centering
    \includegraphics[width=1\linewidth]{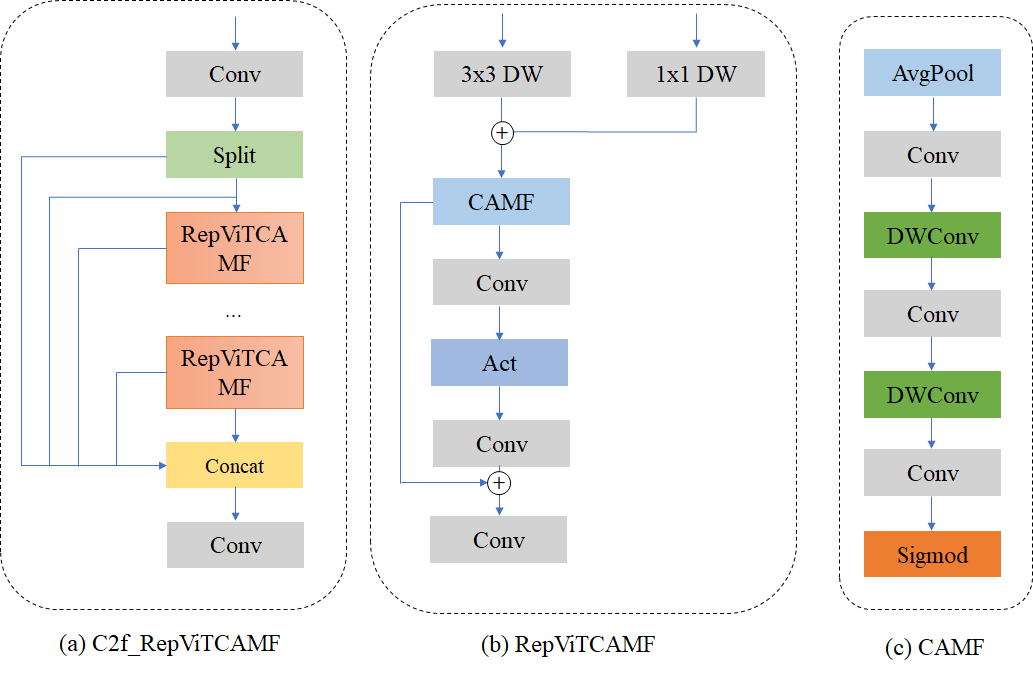}
    \caption{Structure Diagram of C2f\_RepViTCAMF Modules}
    \label{C2f_RepViTCAMF}
\end{figure}

The C2f\_RepViTCAMF module is designed to enhance the model's ability to detect small pulmonary nodules with both high accuracy and efficiency. In medical imaging, particularly for detecting pulmonary nodules, small and fine-grained features are often difficult to capture. This module addresses these challenges by improving the feature extraction and context fusion processes, enabling the model to accurately detect small targets while maintaining computational efficiency. It leverages the improved RepViT module and introduces the Contextual Attention with Multi-scale Feature Fusion (CAMF) module. These advancements are crucial for achieving high performance in real-time medical image analysis, allowing the model to efficiently capture fine-grained features, integrate multi-scale information, and process medical images swiftly. The improved network architecture is shown in Figure \ref{C2f_RepViTCAMF}.

The RepViT module was designed with the goal of achieving a lightweight architecture without compromising performance. It addresses the problem of excessive computational and memory overhead, especially in medical image analysis tasks where high accuracy and efficiency are crucial. To achieve this, RepViT uses structural reparameterization, where skip connections are learned during the training phase but removed during inference, reducing computational load. Additionally, the RepViT module separates spatial and channel processing through two distinct mixers, enhancing the model's ability to handle both types of information efficiently. However, RepViT still faces challenges when dealing with small and complex objects, such as pulmonary nodules, which are often located in fine-grained regions of the lung. This is where the RepViTCAMF module comes into play.

The introduction of the RepViTCAMF module significantly improves upon the original RepViT by incorporating multi-scale context enhancement operations. These operations tackle the challenge of detecting small, dispersed targets in complex medical images. The module leverages depthwise separable convolutions, which reduce the computational complexity by decoupling spatial and channel convolutions while still maintaining effective feature extraction capabilities. This allows the model to focus on small objects without the risk of overloading the system with excessive computation. Additionally, the multi-scale context capturing module in RepViTCAMF enhances the model's ability to simultaneously capture local and global features. This is particularly important for small pulmonary nodules, as the model needs to integrate information from both local image details and broader contextual patterns to accurately detect and classify these lesions.

Each module in the C2f\_RepViTCAMF architecture contributes to the overall model's enhanced detection capability. The RepViT module serves as the foundation, providing a lightweight design optimized for efficiency. Its structural reparameterization and separation of spatial and channel information processing ensure that the model remains computationally efficient without sacrificing performance. The RepViTCAMF module builds upon this foundation by introducing context-aware operations that boost the model's ability to handle small pulmonary nodules. The depthwise separable convolutions reduce unnecessary computational overhead, enabling the model to focus on fine-grained feature extraction. The multi-scale context capturing module allows the model to operate at different scales, providing a comprehensive view of the image and improving its ability to identify small objects. Furthermore, the context fusion operation in RepViTCAMF merges both channel and spatial information to create a more comprehensive feature representation. This fusion strengthens the model's ability to detect small and dispersed targets, improving both localization and classification accuracy. By improving feature extraction, enhancing context awareness, and maintaining a lightweight design, the C2f\_RepViTCAMF architecture significantly contributes to the model’s ability to detect small pulmonary nodules with high precision, making it well-suited for fast and accurate real-time medical imaging analysis.

\subsection{CPYOLO Neck}

The CPYOLO Neck module aims to overcome the difficulties of detecting objects at different scales and improve the fusion of multi-scale features. In the context of pulmonary nodule detection, small objects often lack clear, distinct features and are surrounded by complex backgrounds, making them hard to identify. To tackle this, the Neck module focuses on effective feature extraction and multi-scale feature fusion. The integration of the Feature Pyramid Network (FPN) enables the model to process features at different resolutions, ensuring both small and large objects are accurately detected. Additionally, the incorporation of the KAN-Bottleneck further optimizes the feature extraction process, particularly in low-contrast or imbalanced datasets, enabling the model to better handle the detection of small targets, such as pulmonary nodules.

\begin{figure}
    \centering
    \includegraphics[width=1\linewidth]{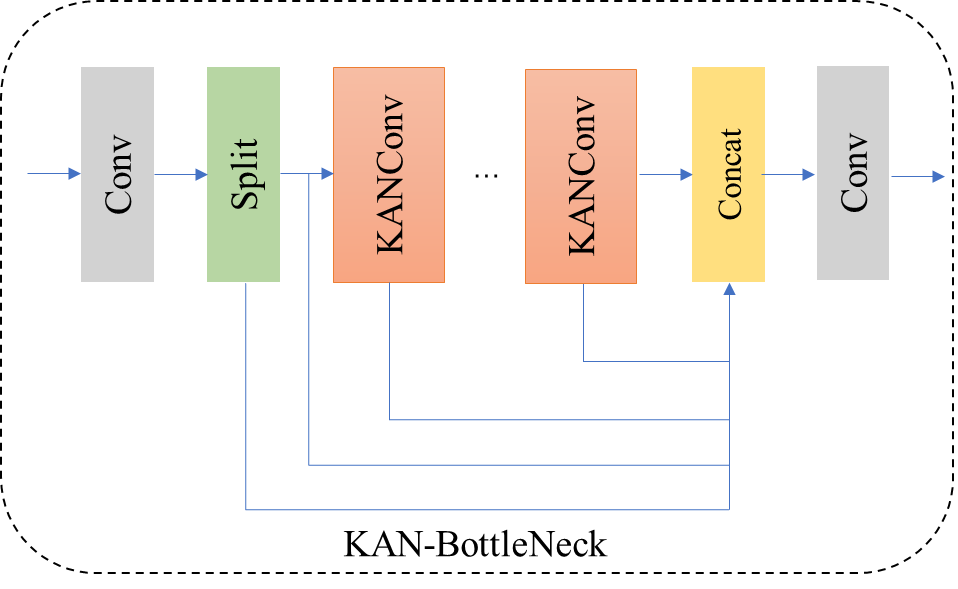}
    \caption{Structure Diagram of KAN\_BottleNeck Module}
    \label{KAN_BottleNeck}
\end{figure}

In addition to FPN, CPYOLO integrates a KAN-Bottleneck to improve the feature extraction process, as shown in Figure\ref{KAN_BottleNeck}. The Bottleneck module is crucial for reducing the dimensionality of feature maps and subsequently restoring them, which reduces computational cost while increasing network depth. The KAN-Bottleneck replaces traditional convolutional layers with the KAN network, which avoids the linear combination of inputs and directly applies nonlinear activation functions to the input data. This design enhances the model's ability to learn complex feature mappings, particularly when dealing with small targets such as pulmonary nodules.

The KAN network's design enables better feature learning through its learnable univariate activation functions and external function summation. This increases its flexibility and efficiency, especially when training samples are limited or the dataset is imbalanced. The mathematical expression for the KAN network is as follows:

\[
f(x) = \sum_{p=1}^{n} \phi_{q,p}(x) \quad \text{(Summation of internal functions)}
\]

where \(\phi_{q,p}(x)\) represents the learnable internal functions, and the summation of external functions is given by:

\[
f(x) = \sum_{q=1}^{n} \Phi_q(x) \quad \text{(Summation of external functions)}
\]

By incorporating KAN-Bottleneck, the CPYOLO model enhances the extraction of refined features, improving detection accuracy for small pulmonary nodules.

Furthermore, CPYOLO integrates two Convolutional Block Attention Modules (CBAM) within the Neck module to further enhance feature extraction. The CBAM modules apply attention mechanisms in both channel and spatial dimensions, focusing on the most relevant features. The channel attention mechanism assigns weights to channels based on their importance, while the spatial attention mechanism highlights critical regions in the input feature maps. These attention mechanisms allow the model to prioritize important features and suppress irrelevant ones.

The CBAM module consists of two sub-modules: the Channel Attention Mechanism and the Spatial Attention Mechanism. The Channel Attention Mechanism operates by applying global average pooling (GAP) and global max pooling (GMP) to the input feature map \( F \), producing two spatial representation vectors. These vectors are processed by a Multi-Layer Perceptron (MLP) to generate the final channel-wise attention weights. The mathematical expression for the channel attention mechanism is as follows:

\[
M_c(F) = \sigma\left( \text{MLP}\left(\text{AvgPool}(F)\right) + \text{MLP}\left(\text{MaxPool}(F)\right) \right) \tag{3.7}
\]

where \( F \) denotes the input feature map, \(\text{AvgPool}(F)\) represents global average pooling, \(\text{MaxPool}(F)\) represents global max pooling, MLP is the multi-layer perceptron, and \(\sigma\) is the Sigmoid activation function.

The Spatial Attention Mechanism highlights important spatial regions in the input feature map. It begins by applying GAP and GMP along the channel dimension to generate two separate feature maps. These maps are concatenated and passed through a convolutional layer with a 7x7 kernel to produce a spatial attention map. This map is then processed by a Sigmoid activation function to generate the final spatial attention feature map. The mathematical expression for the spatial attention mechanism is as follows:

\[
M_s(F) = \sigma \left( f_{7 \times 7} \left( [\text{AvgPool}(F); \text{MaxPool}(F)] \right) \right) \tag{3.9}
\]

In this equation, \(\text{AvgPool}(F)\) and \(\text{MaxPool}(F)\) represent global average pooling and global max pooling operations, respectively, and \(f_{7 \times 7}\) is the convolutional kernel applied to the concatenated pooled features. 

By combining the KAN-Bottleneck and CBAM modules, CPYOLO effectively enhances its ability to focus on important features and suppress irrelevant ones, leading to improved detection performance, especially in tasks such as pulmonary nodule detection.

\subsection{Transfer learning}

In pulmonary nodule detection, ensuring the generalization and stability of the model is crucial, which requires extensive and diverse training data. However, data acquisition in real-world settings is challenging, and manual annotation is time-consuming and labor-intensive. Transfer learning allows the reuse of a pre-trained model from one domain to another, thereby reducing the costs associated with data acquisition and labeling. The central idea of transfer learning is to leverage the labeled data from the source domain, and through algorithmic development, maximize the use of this knowledge. The process involves identifying similarities between the source and target domains to effectively transfer the knowledge from the source domain to the target domain \cite{pan2009survey}. 

To improve the effectiveness of transfer learning, this study proposes enhancements for both the source and target domains. For the source domain, the MS COCO dataset is selected, which differs from datasets such as PASCAL VOC \cite{everingham2015pascal}, as it contains a diverse range of 80 categories. Using MS COCO as the source domain enables the model to learn general, low-level features that can be applied across different domains. As a result, fine-tuning the target domain model with the source domain model allows for the effective transfer of low-level features, thus enhancing the model's ability to generalize. For the target domain, pulmonary nodule images taken in low-contrast or nighttime conditions are first passed through an image enhancement model, such as Zero-DCE, to increase their similarity to normal or daytime pulmonary nodule images, thereby facilitating positive transfer. The model is then fine-tuned on the source domain model, maximizing the use of the knowledge contained in the source domain model.

\section{Experiments}

\subsection{Datasets}

The LUNA16 dataset, used in this study, is a publicly available lung CT imaging dataset derived from the LIDC-IDRI dataset. It was originally part of the 2016 LUNA16 challenge, which focused on the detection of pulmonary nodules. The dataset consists of 888 cases with a total of 36,378 nodules, of which only those larger than 3 mm in size were selected for further analysis. Among these, 1,186 nodules were annotated as positive (valid) nodules. The nodule locations and associated information were meticulously annotated by a team of four radiologists. For the purpose of this study, the three-dimensional CT data were converted into two-dimensional slices to facilitate the analysis of lung nodules.

The LIDC-IDRI dataset is a publicly available lung imaging database jointly released by the National Cancer Institute (NCI) and the National Institutes of Health (NIH) in the United States. It contains 1,010 cases with a total of 1,018 full lung CT scan sequences. Each scan set includes between 200 and 500 CT images. Four radiologists used a "double reading" method to annotate pulmonary nodules. First, each radiologist independently reviewed the CT images and marked the nodule locations. Then, the annotations made by each radiologist were re-evaluated by the other three radiologists to ensure the nodule annotations were as accurate as possible. The annotations were stored in corresponding XML files, and for pulmonary nodules with a diameter greater than 3 mm, the radiologists marked the nodule's contour coordinates and provided subjective semantic feature ratings. These ratings primarily included lobulation degree, calcification degree, spiculation, malignancy, and other characteristics.

\subsection{Data Pre-process}

We propose a method that combines threshold-based segmentation with morphological operations for precise lung tissue extraction. This method minimizes the interference from non-lung regions, such as bones and blood vessels, in the segmentation of the lung parenchyma. First, we automatically binarize the original lung CT images using Otsu's method, which calculates an optimal threshold based on the image's intensity distribution. This effectively separates the lung parenchyma from surrounding tissues. Next, we perform an area-opening operation, which removes small noise or bubble-like regions within the segmented lung parenchyma. Next, we perform another area-opening operation to eliminate any remaining small connected components. This refines the segmentation. Finally, we perform a pixel-wise AND operation on the lung parenchyma mask and the original CT image. This operation guarantees that only lung tissue is kept, excluding all other regions. This method guarantees high-precision segmentation. This ensures reliable data for subsequent medical analysis and reduces external interference.

\subsection{Implementation Details}

We conducted the experiments on an Ubuntu 22.04 LTS system with high-performance hardware for efficient deep learning. The setup includes an Intel Core i9-13900K processor with 24 cores and 32 threads, providing robust parallel processing for large-scale computations. The 128GB of RAM ensures seamless handling of extensive datasets, especially those involving image and video processing. The NVIDIA RTX 4090 with 24GB of VRAM provides unrivaled computational power, essential for training sophisticated models like Convolutional Neural Networks (CNNs). A 1TB NVMe SSD ensures fast data read and write speeds, reducing I/O bottlenecks during training. For the software side, Python 3.8 or 3.9 was selected to maintain compatibility with deep learning libraries. For the same reasons, PyTorch 1.13 or 2.0 was chosen. The CUDA 11.7 toolkit was used to leverage parallel computing resources, accelerating model training. The training process was executed over 100 to 150 epochs, with batch sizes of 16 or 32 to optimize GPU resource usage. The learning rate was set at 0.001 or 0.0005 to prevent unstable gradient updates. We used the Adam or AdamW optimizer for faster convergence. We set the weight decay to 0.0001 to mitigate overfitting and improve model robustness.

\subsection{Results}
To demonstrate the superiority of the proposed algorithm in the task of pulmonary nodule detection, we compare it not only with classical object detection models but also with more modern models. The experimental results on the LUNA16 dataset are shown in Table \ref{exp_LUNA16}, and the results on the LIDC-IDRI dataset are shown in Table\ref{exp_LIDC}. In terms of overall detection performance, the proposed algorithm outperforms the comparison models across three key metrics: precision, recall, and mean average precision (mAP). Notably, both SSD (Single Shot MultiBox Detector) and YOLOv7 exhibit relatively poor performance on these metrics, ranking at the lowest levels. This indicates that the proposed algorithm not only enhances detection accuracy but also demonstrates high robustness in complex tasks.

\begin{table*}[h]
\centering
\caption{Experimental Results of Detection Performance Comparison of Different Models on LUNA16}
\label{exp_LUNA16}
\begin{tabular}{lcccccc}
\hline
\textbf{Model} & \textbf{Precision (\%)} & \textbf{Recall (\%)} & \textbf{Map50 (\%)} & \textbf{Map50-95 (\%)} & \textbf{Fps (f/s)} & \textbf{Weights (Mb)} \\ \hline
Faster R-CNN & 83.8±0.08 & 81.4±0.08 & 87.2±0.05 & 42.4±0.09 & 89.4 & 26.4 \\ 
SSD & 82.7±0.09 & 79.0±0.08 & 85.2±0.08 & 40.8±0.08 & 97.7 & 22.9 \\ 
YOLOv5 & 83.6±0.08 & 81.1±0.09 & 87.0±0.08 & 42.0±0.05 & 127.3 & 2.65 \\ 
YOLOv6 & 84.8±0.05 & 82.1±0.08 & 87.7±0.12 & 43.6±0.09 & 113.3 & 8.94 \\ 
YOLOv7 & 83.3±0.05 & 80.2±0.05 & 86.2±0.08 & 41.7±0.08 & 95.6 & 70.3 \\
YOLOv8 & 86.3±0.12 & 82.8±0.08 & 88.3±0.12 & 45.0±0.12 & 118.2 & 4.76 \\ 
Gold-YOLO & 87.6±0.08 & 85.3±0.09 & 89.5±0.08 & 46.7±0.05 & 102.2 & 22.6 \\ 
MS-YOLO & 86.7±0.05 & 83.7±0.08 & 88.9±0.05 & 45.8±0.08 & 89.6 & 24.1 \\ 
Ghostnetv2 & 87.1±0.08 & 84.1±0.05 & 89.2±0.08 & 46.1±0.08 & 122.1 & 4.03 \\ 
RT-DETR & 88.6±0.05 & 88.2±0.09 & 90.2±0.08 & 47.6±0.05 & 90.4 & 62.1 \\ 
YOLOv9 & 89.6±0.09 & 89.1±0.09 & 90.8±0.05 & 49.1±0.05 & 68.1 & 23.2 \\ 
CPYOLO & 90.8±0.05 & 89.8±0.05 & 92.7±0.08 & 50.6±0.09 & 132.4 & 2.69 \\ \hline
\end{tabular}
\end{table*}

The algorithm proposed in this paper demonstrates excellent performance in terms of frames per second (FPS), showcasing significant advantages in inference speed. Additionally, the model weights of the proposed algorithm are smaller, which is crucial for practical applications. Smaller model weights not only reduce storage and computational overhead but also accelerate the inference process. Compared to the baseline models, the proposed algorithm achieves a significant improvement in FPS, reflecting substantial optimization in inference efficiency. This optimization is primarily attributed to the improved RepViT module in the model, which enhances detection speed through efficient feature extraction and reduction of redundant computations, ensuring high precision without sacrificing inference efficiency.

Although YOLOv9 outperforms YOLOv8 in detection accuracy, its weight reaches 23.2M, several times that of YOLOv8, and its model structure is more complex. While YOLOv9 improves detection performance, its complexity does not lead to a significant performance boost in the lung nodule detection task. This experimental result clearly indicates that, in medical detection tasks such as lung nodule detection, pursuing model complexity alone does not significantly increase detection performance. In fact, optimizing the network structure and training methods tailored to specific tasks can better enhance model performance. In lung nodule detection, the model not only needs high detection accuracy but also requires strong speed and efficiency to meet the demands for fast inference and efficient computation in medical applications. Therefore, relying solely on complex model designs without considering task characteristics will not result in optimal detection performance.

In terms of feature extraction capabilities, classic models such as Faster R-CNN, SSD, and YOLOv5 exhibit relatively weaker feature extraction abilities compared to the improved model proposed in this paper. For instance, in Faster R-CNN, the basic convolutional neural network (e.g., VGG-16) loses some spatial information due to several pooling layers, leading to lower resolution and an inability to effectively capture small objects. The SSD model directly predicts on feature maps at different levels, but its high-level feature maps lose resolution after multiple down-sampling operations, resulting in poor detection performance for small objects. Although YOLOv5 is optimized with a feature pyramid network (FPN), it remains relatively simple compared to more complex multi-stage networks. Thus, it struggles with multi-scale feature fusion and capturing fine details, further affecting small object detection performance.

In contrast, newer models like YOLOv8 and YOLOv9 employ deeper network layers and powerful attention mechanisms, enhancing feature extraction capabilities. However, this complexity increases computational overhead and reduces inference speed. In medical applications, especially in tasks like lung nodule detection, computational complexity and inference speed become crucial factors for practical deployment. Although YOLOv8 and YOLOv9 offer improvements in accuracy, their higher computational complexity and larger model sizes make them less efficient for resource-constrained devices. Therefore, despite their accuracy improvements, the higher computational complexity and larger model sizes of YOLOv8 and YOLOv9 make them less efficient for deployment in resource-limited environments.

\begin{table*}[h]
\centering
\caption{Experimental Results of Detection Performance Comparison of Different Models on LIDC-IDRI}
\label{exp_LIDC}
\begin{tabular}{lcccccc}
\hline
\textbf{Model} & \textbf{Precision (\%)} & \textbf{Recall (\%)} & \textbf{Map50 (\%)} & \textbf{Map50-95 (\%)} & \textbf{Fps (f/s)} & \textbf{Weights (Mb)} \\ \hline
CenterNet & 88.8±0.13 & 86.4±0.13 & 90.2±0.10 & 47.4±0.14 & 94.4 & 29.4 \\ 
SSD & 87.7±0.14 & 84.0±0.13 & 89.2±0.13 & 45.8±0.13 & 102.7 & 27.9 \\ 
YOLOv5 & 88.6±0.13 & 86.1±0.14 & 91.0±0.13 & 47.0±0.10 & 132.3 & 3.15 \\ 
YOLOv6 & 89.8±0.10 & 87.1±0.13 & 92.7±0.17 & 48.6±0.14 & 118.3 & 10.44 \\ 

YOLOv8 & 91.3±0.17 & 87.8±0.13 & 93.3±0.17 & 50.0±0.17 & 123.2 & 7.26 \\ 

MS-YOLO & 91.7±0.10 & 88.7±0.13 & 93.9±0.10 & 50.8±0.13 & 94.6 & 29.1 \\ 
Ghostnetv2 & 92.1±0.13 & 89.1±0.10 & 94.2±0.13 & 51.1±0.13 & 127.1 & 6.53 \\ 
EfficientDet-Do & 93.6±0.10 & 93.2±0.14 & 95.2±0.13 & 52.6±0.10 & 95.4 & 67.1 \\ 
YOLOv9 & 94.6±0.14 & 94.1±0.14 & 95.8±0.10 & 54.1±0.10 & 73.1 & 28.2 \\ 
CPYOLO & 95.8±0.10 & 94.8±0.10 & 97.7±0.13 & 55.6±0.14 & 137.4 & 3.19 \\ \hline
\end{tabular}
\end{table*}

The experimental results in the table\ref{exp_LIDC} demonstrate the performance of various detection models on the LIDC-IDRI dataset, showing a comparison across precision, recall, mAP50, mAP50-95, FPS, and model weights. CPYOLO outperforms the other models, achieving the highest precision (95.8\%) and recall (94.8\%), indicating its strong accuracy in detecting lung nodules while maintaining a high level of true positive detection. It also leads in mAP50 (97.7\%) and mAP50-95 (55.6\%), showcasing its effectiveness across multiple IoU thresholds. Additionally, CPYOLO delivers the highest FPS (137.4 f/s), which is crucial for real-time inference in medical applications. However, its relatively larger model size (3.19MB) might be a concern for devices with limited storage. YOLOv9 provides competitive detection accuracy with 95.8\% mAP50 and 54.1\% mAP50-95, but its FPS is lower (73.1 f/s), which could affect its practical deployment in time-sensitive environments. Models like YOLOv5 achieve a good balance between speed and model size, with the smallest weight (3.15MB), but its accuracy is slightly lower than the more complex models. Meanwhile, YOLOv8 strikes a balance with good accuracy (93.3\% mAP50) and inference speed (123.2 f/s), making it a viable choice for real-time tasks. Overall, the results indicate that, in medical detection tasks like lung nodule detection, optimizing both detection accuracy and inference speed is key, with models like CPYOLO offering the best trade-off for efficiency and performance, while models like YOLOv8 and YOLOv5 are more suitable for environments with resource constraints.

CPYOLO integrates the C2f\_RepViTCAMF, MSCAF, and KAN modules to enhance feature extraction capabilities and inference speed for lung nodule detection tasks. The C2f\_RepViTCAMF module effectively captures long-range pixel contextual relationships, improving feature extraction efficiency and speed for small objects. The MSCAF module further strengthens feature extraction by fully leveraging both channel and spatial information, thereby improving overall detection accuracy. The KAN module enables the model to adaptively adjust activation functions based on data characteristics, capturing more complex data patterns and subtle differences, which enhances feature extraction. The synergy of these modules not only optimizes the model’s training and inference speed but also ensures high detection performance, particularly demonstrating outstanding efficiency in real-time detection tasks. Experiments on the LIDC-IDRI dataset show that the proposed model achieves higher FPS compared to most baseline models, reflecting its efficiency in practical applications. The smaller model weights and optimized feature extraction mechanism ensure high detection performance while avoiding significant computational overhead, making the algorithm particularly suitable for deployment in resource-constrained environments, which is crucial for real-world medical applications. Although YOLOv9 outperforms YOLOv8 in detection accuracy, given that this model is primarily applied to medical scenarios, particularly lung nodule detection, we emphasize that optimizing speed, efficiency, and computational resources is equally important. Therefore, model evaluation should not rely solely on detection accuracy but also consider training and inference speed as well as computational efficiency. The weight of YOLOv9 is 21 times that of YOLOv8, with a higher parameter count and computational complexity, making it unsuitable for resource-constrained devices. In contrast, YOLOv8, with its lower computational complexity and better stability, is more suitable for deployment in practical medical applications. Ultimately, the improved model based on YOLOv8 not only increases detection accuracy but also reduces computational complexity, significantly enhancing efficiency and practicality.

\subsection{Ablation Study}

As shown in Table 2, through a series of targeted optimizations, the improved model proposed in this paper has achieved significant improvements across all evaluation metrics. Firstly, the RepViT module was specially modified and integrated with the C2f module. The optimized model has greatly enhanced its feature extraction and representation capabilities, allowing it to more accurately capture the subtle characteristics of lung nodules, thereby improving key metrics such as precision and recall. In addition, the Neck section of the model was also improved, replacing the original feature fusion component. The new Neck design more effectively handles multi-scale features, demonstrating stronger capabilities in detecting nodules of different sizes, which led to an approximately 3\% improvement in detection performance. Notably, the MSCAF module, through the introduction of multi-scale features and channel attention mechanisms, successfully addressed the issue of scale variations in lung nodules, enhancing detection accuracy for nodules of various sizes. 

The improved model outperforms the original model significantly. It achieved a 4.5\%±0.13 persen increase in precision, a 7.0\%±0.09 persen increase in recall, a 4.4 persen±0.14 improvement in mAP50, and a 5.7 persen±0.15 increase in mAP50-95. These significant improvements demonstrate that the model enhances detection accuracy and sensitivity while maintaining high inference efficiency and low computational complexity, showcasing outstanding performance. The improved model is particularly useful in the lung nodule detection task.

\begin{table*}[h]
\centering
\caption{Ablation Experiment Results}
\begin{tabular}{ccccccc}
\hline
\textbf{C2f\_RepViTCAMF} & \textbf{MSCAF} & \textbf{KAN-Bottleneck} & \textbf{Precision (\%)} & \textbf{Recall (\%)} & \textbf{mAP50 (\%)} & \textbf{mAP50-95 (\%)} \\ \hline
— & — & — & 87.3±0.12 & 83.8±0.08 & 89.3±0.12 & 46.0±0.12 \\ 
√ & — & — & 88.9±0.08 & 86.5±0.08 & 92.1±0.12 & 49.6±0.08 \\ 
— & √ & — & 90.4±0.12 & 88.2±0.12 & 92.8±0.08 & 50.4±0.16 \\ 
√ & √ & — & 90.6±0.05 & 88.4±0.12 & 93.0±0.17 & 50.6±0.06 \\ 
√ & √ & √ & 91.8±0.05 & 90.8±0.05 & 93.7±0.08 & 51.7±0.09 \\ \hline
\end{tabular}
\end{table*}

\subsection{Visualization}
\begin{figure}
    \centering
    \includegraphics[width=1\linewidth]{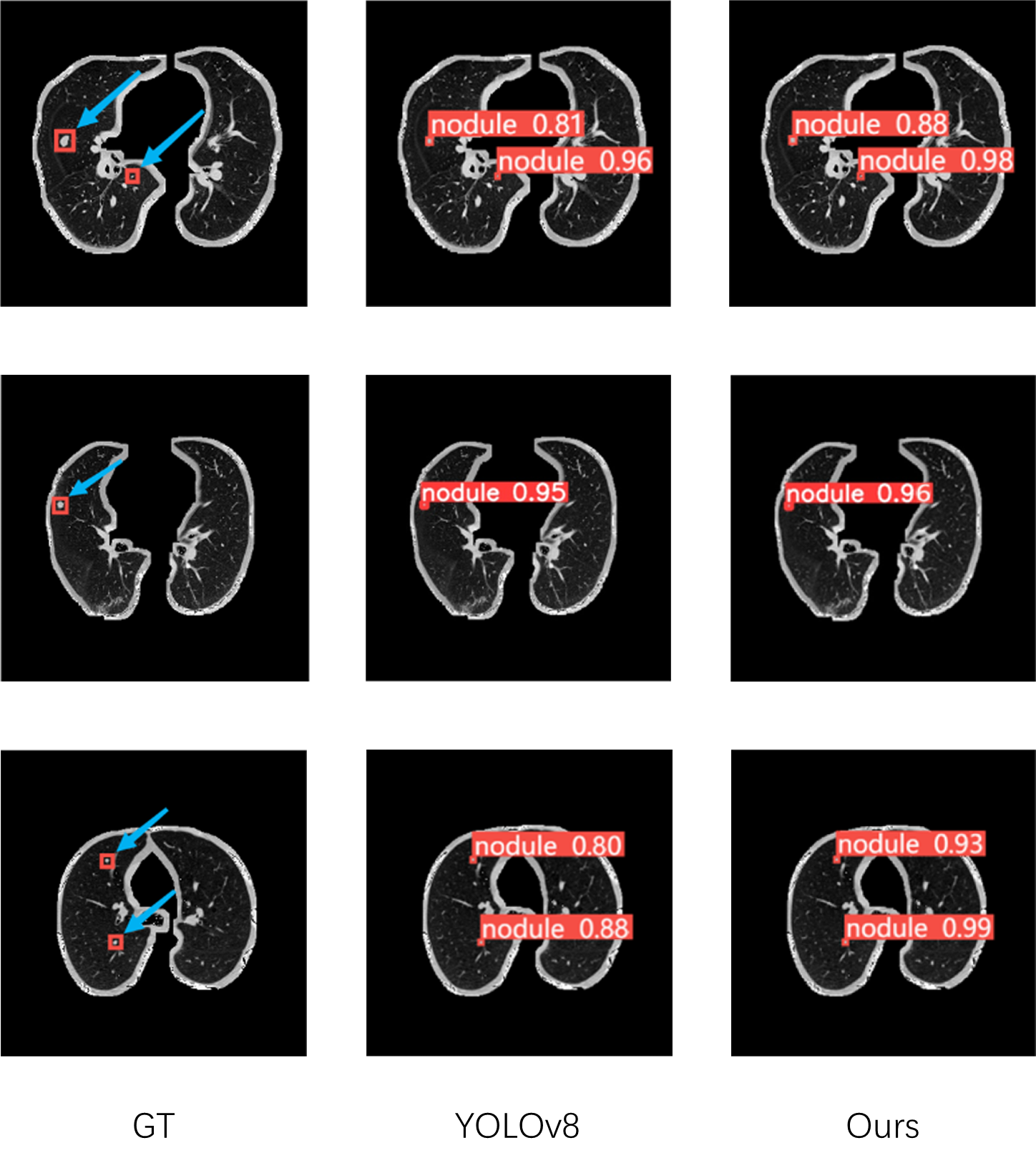}
    \caption{Visual comparison of lung nodule detection results on the LUNA16}
    \label{Visualization}
\end{figure}

Lung parenchyma segmentation significantly enhances detection accuracy by removing background noise. This process involves image normalization, followed by the application of the K-means clustering algorithm to differentiate lung tissue from the background. Additionally, morphological operations such as erosion and dilation are employed to refine the lung mask. As illustrated in Figure \ref{Visualization}, the segmented image shows a marked improvement in performance compared to the original data. In the detection of multiple nodules, the segmentation method achieves an accuracy of 98\%, effectively reducing both false positives and false negatives. Overall, the application of lung parenchyma segmentation across all models substantially improves detection accuracy, highlighting its critical role in complex image analysis.

\section{Conclusion}
The integration of the Internet of Things (IoT) into medical imaging systems presents a transformative opportunity for enhancing real-time diagnostics and patient monitoring. Building on this foundation, we have introduced CPYOLO, a novel method for detecting pulmonary nodules in CT images, specifically designed to address the challenges of multi-scale nodule detection and ensure accurate identification. By optimizing the C2f (Cascaded Feature Fusion) module within the backbone network through the C2f\_RepViTCAMF module, our model significantly enhances the detection accuracy of small pulmonary nodules while maintaining a lightweight design. This improvement boosts the feature representation capability, enabling the model to capture the subtle features of small nodules more effectively. Furthermore, our optimized feature fusion module adeptly handles the substantial scale variation of pulmonary nodules across different cases, overcoming the limitations of fixed-scale detection and thereby improving the model's accuracy for multi-scale nodules. The incorporation of the Kernelized Attention Network (KAN) leverages an adaptive attention mechanism, further enhancing the model's ability to extract subtle features of small nodules and overall detection accuracy. Empirical results demonstrate that CPYOLO not only excels in detecting small nodules amidst complex backgrounds but also surpasses existing benchmarks on the public LUNA16 dataset.

Despite these advancements, the proposed algorithm exhibits limitations when processing pulmonary CT images with high background noise, where detection accuracy for small or blurry nodules diminishes. Future research will focus on enhancing the model's noise resistance by introducing large-scale models with deeper network structures and more parameters, thereby improving noise suppression capabilities. Additionally, integrating multimodal data, such as combining clinical medical reports with pulmonary CT images, will further enhance the model’s performance. This multimodal fusion~\cite{yunfan2025mitigating} will enable more robust feature extraction and reduce the impact of noise on detection performance, ensuring greater stability and precision in clinical settings. In summary, CPYOLO represents a significant advancement in medical image detection, particularly for pulmonary nodules, and its integration with IoT-driven medical systems holds the potential to revolutionize future medical image processing tasks, leading to more accurate and efficient diagnostic tools in healthcare.

In future research, we plan to enhance CPYOLO's applicability by integrating multimodal data, such as patient medical histories, clinical reports, and other imaging modalities like MRI and PET scans, alongside pulmonary CT images. This fusion of diverse data sources can significantly improve the model's robustness and diagnostic accuracy. By incorporating clinical data, the model gains valuable contextual insights, such as pre-existing conditions, comorbidities, and treatment history, which can aid in distinguishing between benign and malignant nodules or identifying atypical patterns that imaging alone may miss. Furthermore, combining different imaging modalities provides complementary information, helping to overcome the limitations of individual methods. For example, MRI and PET scans offer better tissue characterization, which, when integrated with CT scans, provides a more comprehensive view of nodule features and tissue boundaries. This multimodal approach will bolster the model's ability to detect small or blurry nodules in noisy environments, thereby improving overall accuracy and reducing false negatives. Ultimately, integrating multimodal data will strengthen feature extraction, enhance the model’s adaptability in various clinical scenarios, and ensure that CPYOLO delivers more precise and personalized diagnostic support, making it a valuable tool in real-time diagnostic systems, particularly those powered by IoT.

\section*{CRediT authorship contribution statement}
Meng Wang is responsible for research design, model architecture design and core algorithm development, and leads the writing of papers. Zi Yang is responsible for data pre-processing and analysis, and participates in model training and tuning, as well as result collation. As the corresponding author, Ruifeng Zhao coordinates the team's work, provides technical guidance, and participates in the review and revision of the paper. Yaoting Jiang provides deep learning suggestions, participates in experimental analysis and discussion, and provides theoretical support for paper writing.

\section*{Declaration of Interest Statement}
The authors declare that they have no known competing financial interests or personal relationships that could have appeared to influence the work reported in this paper.

\section*{Data availability}
Data will be made available on request.


\bibliographystyle{model1-num-names}
\bibliography{cas-refs}


\end{CJK}
\end{sloppypar}
\end{document}